\documentclass[12pt,a4paper]{article}
\usepackage[utf8]{inputenc}
\usepackage[english]{babel}
\usepackage{dmstyle}
\usepackage[normalem]{ulem}
\usepackage{cite}

\def \LambdaNP{\Lambda_{_{\rm NP}}}
\def \varsig {\varsigma}
\def \PKL {{\rm P}}

\begin{document}

\title{Extremal isolated horizons with $\Lambda$ \\
and the related unique type D black holes}

\author{David Matejov, Ji\v{r}\'{\i} Podolsk\'{y} 	\\[4mm]
	{\small Charles University, Faculty of Mathematics and Physics,} \\
	{\small Institute of Theoretical Physics, V~Hole\v{s}ovi\v{c}k\'ach~2, 180~00 Prague 8, Czech Republic
	} \\[3mm]
    {\small  E-mail: \texttt{d.matejov@gmail.com, podolsky@mbox.troja.mff.cuni.cz}}\\}

\maketitle

\begin{abstract}

We extend our previous work in which we derived the most general form of an induced metric describing the geometry of an axially symmetric extremal isolated horizon (EIH) in asymptotically flat spacetime. Here we generalize it to EIHs in asymptotically (anti-)de Sitter spacetime. The resulting metric conveniently forms a 6-parameter family which, in addition to a cosmological constant~$\Lambda$, depends on the area of the horizon, total electric and magnetic charges, and two deficit angles representing conical singularities at poles. Such a metric is consistent with results obtained in the context of near-horizon geometries. Moreover, we study extremal horizons of all black holes within the class of Plebański-Demiański exact (electro)vacuum spacetimes of the algebraic type~D. In an important special case of non-accelerating black holes, that is the famous Kerr-Newman-NUT-(A)dS metric, we were able to identify the corresponding extremal horizons, including their position and geometry, and find explicit relations between the physical parameters of the metric and the geometrical parameters of the EIHs.

\end{abstract}

\vfil\noindent
PACS class:  04.20.Jb, 04.40.Nr, 04.70.Bw, 04.70.Dy, 04.20.-q, 97.60.Lf



\newpage

\section{Introduction}

In the present article, we continue our investigation of axisymmetric extremal isolated horizons admitting a non-zero value of the cosmological constant~$\Lambda$.

As we have already pointed out in our previous work \cite{MatejovPodolsky}, the concept of an isolated horizon has many interesting and advantageous features. Above all, it may serve as a model describing a black hole in equilibrium with its neighborhood (its accretion disk, an external electromagnetic field, etc.), purely (quasi-)locally. This can be very useful in theoretical research as well as in various applications in numerical relativity or related astrophysical studies \cite{Ashtekar2004-review, Krishnan-2012, Ashtekar2004-multipolemoments, Gurlebeck2015}. Among  significant recent discoveries let us mention the general proof of the Meissner effect for black holes \cite{Scholtz2017, Scholtz2018}. Our work continues along this direction. More specifically, we rigorously analyse the \emph{uniqueness} of the extremal black hole horizons.

It has been previously shown \cite{Lewandowski-Pawlowski, BukLewandowski} that when a black hole becomes extremal (by increasing its rotation, for example), it exhibits behaviour leading to its very special properties that do not depend on the surrounding environment. One of these properties is the uniqueness of the induced metric on the horizon slices of constant time. Here we extend our previous investigations and results \cite{MatejovPodolsky} to the case when the black hole is situated in asymptotically (anti-) de Sitter spacetime with a non-zero cosmological constant. We systematically derive the induced metric of the extremal horizon using the Newman-Penrose (NP) formalism, pointing out differences between the ${\Lambda=0}$ and ${\Lambda\neq 0}$ cases. We also compare our general result with the analogous one previously obtained in \cite{KunduriLucietti, KunduriLucietti2009a, LiLucietti2013, KunduriLucietti2009b}. We discuss the advantages of our approach, leading to a result which --- by its simple and elegant form --- allows also direct interpretation of the obtained integration constants. In particular, we find explicit relations between geometrical parameters of the EIHs and physical parameters of the Kerr-Newman-NUT-(anti-)de Sitter solution contained in the Plebański-Demiański class of metrics
\cite{Plebanski1976,Debever1971, Stephani, Griffiths2009, Griffiths2005, GriffithsPodolsky2006, PodolskyGriffiths2006, PodolskyVratny2021}.

Let us summarize structure of this paper. In Sec.~2, we review the necessary notation and
basic definitions concerning isolated horizons. In Sec.~3 we specialize on extremal isolated horizons with non-zero cosmological constant $\Lambda$, and we explicitly solve the constraint equations for a function describing the horizon geometry. We also compare our result with the analogous result already known in  literature.
In Sec.~4 we investigate the horizon geometry of the most general type D black hole in a Plebański-Demiański family of exact spacetimes. Then we restrict our attention to non-accelerating black holes, that is the well-known Kerr-Newman-NUT-(A)dS spacetime. In the final Sec.~5 we show that such an extremal horizon has geometry identical to the one derived for a generic EIH in Sec.~2, and we also provide explicit relations between the parameters of both solutions.
Appendix~A contains a discussion of the number and character of possible extremal horizons in the Kerr-Newman-NUT-(A)dS spacetime.

\newpage

\section{Preliminaries}

Here we consider EIHs with a non-zero cosmological constant ${\Lambda \neq 0}$. In our convention of metric ${(+---)}$, the Einstein equations read
\begin{align}
R_{ab} - \tfrac{1}{2} R\, g_{ab} + \Lambda\, g_{ab} = - 8\pi\, T_{ab}.
\end{align}
In the Newman-Penrose (NP) formalism\footnote{For its summary see our previous work \cite{MatejovPodolsky}.}, the equations are reduced to a relation between the trace-free part of the Ricci tensor and corresponding tetrad projections of the energy-momentum tensor. In electrovacuum spacetimes with $\Lambda$, this relation is simply
\begin{align}
\Phi_{ab} = 2 \,\phi_a  \bar{\phi}_b,
\end{align}
where $\phi_a$ are tetrad projections of the electromagnetic field tensor $F_{ab}$.

Further we define $\ham$ to be an \emph{isolated horizon} with a cross section $\mathcal{K}$. The null generator of $\ham$ coincides with the null vector $l^a$ of the NP tetrad on $\ham$, while the vectors $m^a, \bar{m}^a$ span the tangent space of $\mathcal{K}$, and $n^a$ is constant on $\ham$.

It turns out that for axially symmetric 2-dimensional manifolds of spherical topology it is useful to introduce adapted coordinates ${(\zeta, \phi)\in [-1,1]\times[0,2\pi)}$ in which its metric has the canonical form \cite{Ashtekar2004-multipolemoments}
\begin{align}  \label{eq:CanonicalMetric}
q_{ab}\, \dd x^a \dd x^b \equiv - R^2\,\Big( \,\frac{1}{f(\zeta)}\, \dd \zeta^2 + f(\zeta)\,\dd \phi^2 \,\Big).
\end{align}
Such metric is characterized by a single \emph{metric function} $f(\zeta)$. We further assume that the function satisfies the generalized regularity conditions at the poles ${\zeta=\pm1}$, namely
\begin{align} \label{eq:RegularityConditions}
f'(\pm 1) = \mp \, 2 \Big( 1 + \frac{\delta_{\pm}}{2\pi}\Big).
\end{align}

A convenient choice of the spatial vector $m^a$ on $\ham$ is
\begin{align} \label{eq:VectorMonIH}
m^a \eqNH \frac{1}{\sqrt{2}\,R} \Big(\sqrt{f(\zeta)}\, \pd^a_{\zeta} + \frac{\ii}{\sqrt{f(\zeta)}}\, \pd^a_{\phi} \Big),
\end{align}
normalized as ${m_a \bar{m}^a=-1}$.
The only independent component of the connection on $\ham$ is then given by the coefficient $a$ defined as
\begin{align} \label{eq:SpinCoefficientA}
a \equiv m_a \bar{\delta}\, \bar{m}^a = \alpha - \bar{\beta} \eqNH - \frac{1}{2\sqrt{2}R} \,\frac{f'(\zeta)}{\sqrt{f(\zeta)}}.
\end{align}
With this choice, $a$ is real on the horizon,  $\bar{a} \eqNH a$, as well as the derivative operator $\delta \equiv m^a \nabla_a \eqNH \bar{\delta}$ acting on a scalar function, namely
\begin{align} \label{eq:OperatorDelta}
\delta \fii \eqNH \frac{1}{\sqrt{2}\,R} \sqrt{f(\zeta)}\, \pd_{\zeta} \fii,
\end{align}
for an arbitrary function $\fii = \fii(\zeta)$.

\subsection{Electromagnetic field and the spin coefficient $\pinp$}

As we have already discussed in \cite{MatejovPodolsky, Scholtz2018}, the tetrad component $\phi_1$ of the electromagnetic field tensor $F_{ab}$ is on the horizon governed by the Maxwell equation which, under an assumption of stationarity ${D \phi_2 \eqNH 0}$, reads
\begin{align}
\delta \phi_1 + 2\pinp\, \phi_1 - \surg \,\phi_2 \eqNH 0.
\end{align}
This equation remains unchanged also in spacetimes with a cosmological constant ${\Lambda \neq 0}$. Similarly, the spin coefficient $\pinp$, which is a subject of a particular NP Ricci identity, remains unaffected. Namely, the equation on the horizon reads
\begin{align} \label{eq:EquationForPi}
\bar{\eth} \pinp \eqNH \surg \,\lambda - \pinp^2.
\end{align}
These two equations can be fully integrated in the axially symmetric \emph{extremal} case ${\surg = 0}$. The explicit solutions in the adapted coordinates are
\begin{align} \label{eq:PhiAndPiNP}
\phi_1 \eqNH \frac{c_{\phi}}{(\zeta + c_{\pi})^2},
\qquad
\pinp \eqNH \sqrt{\frac{f}{2}} \,\frac{1}{R\,(\zeta + c_{\pi})},
\end{align}
in which the integration constants $c_\phi, c_\pi$ depend only on intrinsic properties of the horizon. It is illustrative to express $c_\phi$ in terms of the \emph{physical electric and magnetic charges},
\begin{align} \label{eq:PhysicalCharges}
Q \equiv Q_E + \ii\, Q_M = \frac{1}{2 \pi} \oint_{\mathcal{K}} \! \phi_1 \; \mathrm{vol}(\mathcal{K})
   = \frac{2 R^2}{c_{\pi}^2 - 1}\, c_{\phi}.
\end{align}
Inverting this relation gives
\begin{align}
c_\phi &= \frac{Q}{ 2 R^2}\,(c_\pi^2 - 1).
\end{align}

Notice that the general arguments which previously led to proof of the Meissner effect \cite{Scholtz2017, Scholtz2018} remain valid as well.

\section{Geometry of horizon sections}

In previous section we argued that the electromagnetic field~$\phi_1$ and the spin coefficient~$\pinp$ are independent of the cosmological constant $\Lambda$. However, this might not be expected for the $\Psi_2$ component of the Weyl tensor, and for the horizon geometry described by the metric function $f(\zeta)$. Indeed, repeating the same arguments as in \cite{MatejovPodolsky} we obtain the constrain equations for $\Psi_2$ and $f(\zeta)$ in the form
\begin{align}
\eth \pinp &\eqNH  - \pinp \pinpb -\Psi_2 - 2\LambdaNP, \nonumber\\
\eth \pinp -\overline{\eth \pinp} &\eqNH 2a^2 - 2 \delta a - 2 \Psi_2 + 2\LambdaNP + 4|\phi_1|^2.
\end{align}
Both equations contain additional terms proportional to the NP quantity $\LambdaNP$, which is related to the scalar curvature by $\LambdaNP = R/24$. Therefore, in electrovacuum spacetimes ${\LambdaNP = \Lambda/6}$. From now on, we will use only the cosmological constant $\Lambda$ to avoid confusion.

Combining these two equations to eliminate $\Psi_2$ and using equation \eqref{eq:EquationForPi} we arrive at
\begin{align} \label{eq:EquationForF}
a^2 - \delta a + 2|\phi_1|^2 - \tfrac{1}{2} \Lambda \eqNH {\textstyle \frac{1}{2}} (\pinp - \pinpb)^2 + a (\pinp + \pinpb).
\end{align}
Further, we employ the definition \eqref{eq:SpinCoefficientA} and the expression for the derivative operator \eqref{eq:OperatorDelta} in the adapted coordinates. After some algebra, the final equation for $f(\zeta)$ reads
\begin{align}
|\zeta + c_{\pi}|^4\, f'' + (2\zeta + c_{\pi} + \bar{c}_{\pi}) |\zeta + c_{\pi}|^2\, f'
  &+ (c_{\pi} - \bar{c}_{\pi})^2\,f + 8 R^2 |c_\phi|^2  \nonumber\\
  &  + 2 \Lambda R^2  (\zeta + \bar{c}_{\pi})^2 (\zeta + c_{\pi})^2 \eqNH 0.
\end{align}
The general solution in terms of the integration constants $c_\pi$ and $c_\phi$ has the form
\begin{align}
f(\zeta) &= \frac{4|c_\phi|^2 R^2(1-\zeta^2)}{(|c_\pi|^2 - 1)\,|\zeta+c_\pi|^2} \nonumber\\
& \quad
-  \Lambda R^2 (1 - \zeta^2)\frac{(|c_\pi|^2 - 1)\big(\zeta^2 + 2(c_\pi + \bar{c}_\pi)\zeta\big) + 3c_\pi^2 \bar{c}_\pi^2 + c_\pi \bar{c}_\pi - \big(\zeta^2 + 2(c_\pi + \bar{c}_\pi)^2\big)}{3(|c_\pi|^2 - 1)\,|\zeta+c_\pi|^2},
\end{align}
where we have applied the boundary conditions at both poles ${f(\pm 1) = 0}$ to fix the integration constants. We also impose our generalized regularity conditions \eqref{eq:RegularityConditions} to find the value of the constant $c_\pi$. We thus obtain
\begin{align}
c_\pi &= \frac{ \delta_- -\delta_+\pm 2\,\ii\,\sqrt{A_0 - A_1 + A_2} }{4\pi + \delta_- +\delta_+ -4\pi |Q|^2 R^{-2} -4\pi \Lambda R^2},
\end{align}
where
\begin{align}
A_0 & \equiv (2\pi +\delta_-)(2\pi +\delta_+) - 4\pi^2 |Q|^4 R^{-4}, \nonumber \\
A_1 & \equiv \tfrac{4}{3}\pi (4\pi + \delta_- + \delta_+) \Lambda R^2 - \tfrac{8}{3}\pi^2 |Q|^2 \Lambda, \\
A_2 & \equiv \tfrac{4}{3}\pi^2 \Lambda^2 R^4. \nonumber
\end{align}
Substitution into the formula for $f(\zeta)$ yields a unique solution. We summarize it in the following theorem, which is generalization of \cite{MatejovPodolsky}.

\begin{theorem} \label{th:MetricIHwithLambda}
Let $(\NH, [l^a])$ be an axially symmetric extremal isolated horizon (EIH) of topology $\spaceS^{\delta_+}_{\delta_-}$ in asymptotically (anti-)de Sitter spacetime. Then the geometry of its spherical sections is described by an induced metric $q_{ab}$ in the form \eqref{eq:CanonicalMetric}, where the \emph{dimensionless} metric function $f(\zeta)$ is given
\begin{align} \label{eq:MetricFunctionLambda}
  f_{\text{EIH}}(\zeta) &= (1-\zeta^2)\,  \frac{ d_0 + d_1 \,\zeta + d_2\,\zeta^2 }{ c_0 + c_1 \,\zeta + c_2\,\zeta^2 },
\end{align}
in which
\begin{align}
d_0 & \equiv (2/\pi)(2\pi+\delta_-)(2\pi+\delta_+)
     + \tfrac{1}{3}\Lambda R^2 \big[ 4\pi (\Lambda R^2-5) - 5 (\delta_-+\delta_+) + q^2 \big], \nonumber\\[1mm]
d_1 & \equiv \tfrac{4}{3}\Lambda R^2 (\delta_--\delta_+),  \nonumber\\[1mm]
d_2 & \equiv \tfrac{1}{3}\Lambda R^2 \big[ 4\pi (1-\Lambda R^2) + (\delta_-+\delta_+) - q^2 \big], \nonumber\\[-1mm]
 &  \label{eq:dici}\\[-1mm]
c_0 & \equiv 4\pi (1-\tfrac{1}{3}\Lambda R^2) + (\delta_-+\delta_+) + q^2, \nonumber\\[1mm]
c_1 & \equiv 2 (\delta_--\delta_+),  \nonumber\\[1mm]
c_2 & \equiv 4\pi (1-\Lambda R^2) + (\delta_-+\delta_+) - q^2, \nonumber
\end{align}
and we have denoted
\begin{align} \label{eq:DEf-g^2}
  q^2 & \equiv 4\pi\, \frac{|Q|^2}{R^2}.
\end{align}
The function $f_{\text{EIH}}(\zeta) $ is unique and depends on 6 real independent parameters, namely
$\delta_+, \delta_-, R, \Lambda$ and $Q \equiv Q_E + \ii\, Q_M$. It is well-behaved, and any of these parameters (except $R$ when ${|Q|\neq 0}$) can be set to zero.
\end{theorem}

This is a fully general and explicit result for (axisymmetric) extremal isolated horizons, expressed in terms of geometrical and physical parameters, namely:
\begin{align}
\Lambda    \quad.....\quad & \hbox{cosmological constant}         \nonumber \,,\\
R          \quad.....\quad & \hbox{radius defined by the horizon area } A=4\pi R^2\nonumber \,,\\
q^2        \quad.....\quad & \hbox{dimensionless elmag charge parameter } q^2 = (4\pi)^2(Q_E^2 + Q_M^2)/A     \nonumber \,,\\
\delta_\pm \quad.....\quad & \hbox{two deficit angles at the horizon poles ${\zeta=\pm 1}$, respectively}  \nonumber \,.
\end{align}

In fact, $\Lambda$ and $R$ are combined into a \emph{single} dimensionless parameter $\Lambda R^2$, so that \emph{all} terms entering the coefficients $d_i$ and $c_i$ are dimensionless. Moreover, ${d_1=\tfrac{2}{3}\Lambda R^2\,c_1}$ and ${d_2=\tfrac{1}{3}\Lambda R^2\,c_2}$.

The metric function has to be positive, ${f(\zeta)>0}$, and non-zero except at the poles where ${f(\zeta=\pm 1)=0}$, which restricts range of the parameters.

\vspace{4mm}

There are \emph{two natural subcases} to consider:

\subsubsection*{The case ${\Lambda=0}$}

In the spacetimes with zero cosmological constant $\Lambda=0$ the metric function \eqref{eq:MetricFunctionLambda} acquires much simpler form. The coefficients \eqref{eq:dici} reduce to
\begin{align}
d_0 & = (2/\pi)(2\pi+\delta_-)(2\pi+\delta_+), & c_0 & = 4\pi + (\delta_-+\delta_+) + q^2, \nonumber\\[1mm]
d_1 & = 0, & c_1 & = 2 (\delta_--\delta_+),\label{eq:dici Lambda=0}\\[1mm]
d_2 & = 0, & c_2 & = 4\pi + (\delta_-+\delta_+) - q^2, \nonumber
\end{align}
so the function $f(\zeta)$ simplifies to
\begin{align} \label{eq:MetricFunctionLambda=0}
  f_{\text{EIH}}(\zeta) &= \frac{2}{\pi}\,
  \frac{(2\pi+\delta_-)(2\pi+\delta_+)(1-\zeta^2)}
  {4\pi(1 + \zeta^2)+ \delta_-(1 + \zeta)^2 +\delta_+(1-\zeta)^2 + q^2 (1-\zeta^2)}.
\end{align}
This is exactly the function derived and analysed in our previous work, see Theorem~1 and Eq.~(65) in \cite{MatejovPodolsky}.

\subsubsection*{Regular axes ${\delta_-=0=\delta_+}$}

In the case when the both poles are regular, the coefficients \eqref{eq:dici} simplify to
\begin{align}
d_0 & = 8\pi
     + \tfrac{1}{3}\Lambda R^2 \big[ 4\pi (\Lambda R^2-5) + q^2 \big], & c_0 & = 4\pi (1-\tfrac{1}{3}\Lambda R^2) + q^2, \nonumber\\[1mm]
d_1 & = 0, & c_1 & = 0, \label{eq:diciRegularAxis} \\[1mm]
d_2 & = \tfrac{1}{3}\Lambda R^2 \big[ 4\pi (1-\Lambda R^2) - q^2 \big], & c_2 & = 4\pi (1-\Lambda R^2) - q^2,  \nonumber
\end{align}
and thus the function \eqref{eq:MetricFunctionLambda} takes the form
\begin{align} \label{eq:MetricFunctionRegularAxis}
  f_{\text{EIH}}(\zeta) &= (1-\zeta^2)\,  \frac{ 2 + \tfrac{1}{3}\Lambda R^2 \big[ (\Lambda R^2-5 + \tfrac{1}{4\pi } q^2)
   + (1-\Lambda R^2 - \tfrac{1}{4\pi } q^2)\,\zeta^2  \big]}
  { (1-\tfrac{1}{3}\Lambda R^2 + \tfrac{1}{4\pi } q^2) + (1-\Lambda R^2 - \tfrac{1}{4\pi } q^2 )\,\zeta^2 }.
\end{align}
As we will show below, a metric function of this form can be identified with an extremal isolated horizon of the \emph{Kerr-Newman-(anti-)de~Sitter} black hole. When we set the electromagnetic charges to zero (${Q_E = 0 = Q_M}$, implying ${q^2=0}$) we obtain
\begin{align} \label{eq:BLMetricFunction}
  f_{\text{EIH}}(\zeta) &= (1-\zeta^2)\,\frac{2 + \tfrac{1}{3}\Lambda R^2 \big[ (\Lambda R^2-5) + (1-\Lambda R^2)\,\zeta^2  \big]}
  { (1-\tfrac{1}{3}\Lambda R^2) + (1-\Lambda R^2)\,\zeta^2 }.
\end{align}
This is the result recently presented by Buk and Lewandowski \cite{BukLewandowski}, with a straightforward identification of the variables ${\zeta \equiv x}$, ${f_{\text{EIH}} \equiv P^2}$.

\subsection{Comparison with the general result by Kunduri and Lucietti}

An analogous result to our Theorem~\ref{th:MetricIHwithLambda} for the geometry of an extremal black hole has been presented by Kunduri and Lucietti in \cite{KunduriLucietti} in the context of near-horizon geometries. This result in general admits conical singularities as well as electromagnetic field and the cosmological constant. However, from the analysis performed in \cite{KunduriLucietti} it is not immediately clear which physical quantity is related to which integration constant. In what follows we will compare our result \eqref{eq:MetricFunctionLambda} with the result (83), or Theorem 4.3, in \cite{KunduriLucietti} for the uncharged case ${e=0=g}$. Such metric of the near-horizon geometry reads
\begin{align} \label{eq:KL-LineElement}
\dd s_{KL}^2 = \Gamma(x) (A_0 r^2 \,\dd v^2 + \dd v\, \dd r) + \frac{\Gamma(x)}{\PKL(x)} \dd x^2 + \frac{\PKL(x)}{\Gamma(x)}(\dd \Phi + k r\, \dd v)^2,
\end{align}
with
\begin{align}
\Gamma(x) &= \frac{k^2}{\beta} + \frac{\beta x^2}{4}, \label{eq:Gamma(x)}\\
\PKL(x) &=  -\frac{\beta \Lambda}{12}\, x^4 + \left( A_0 -  \frac{2 \Lambda k^2}{\beta} \right)x^2 + c_1\, x - \frac{4 k^2}{\beta^2} \left( A_0 - \frac{\Lambda k^2}{\beta}\right).
\label{eq:PKL(x)}
\end{align}
When we set ${\dd r = 0}$ the metric degenerates if and only if ${r = 0}$. The horizon, which is a null hypersurface, is therefore located at ${r=0}$. Then the induced metric of a horizon section is
\begin{align} \label{eq:KL-inducedLineElement}
g|_\mathcal{K} = \frac{\Gamma(x)}{\PKL(x)}\, \dd x^2 + \frac{\PKL(x)}{\Gamma(x)}\,\dd \Phi^2,
\end{align}
The poles and the range of the coordinate $x$ are determined by possible roots of the polynomial $\PKL(x)$ such that ${\PKL(x_+) = 0 = \PKL(x_-)}$.

\subsubsection{The case ${\Lambda = 0}$}

For simplicity, let us first assume that ${\Lambda = 0}$. The function $\Gamma(x)$ remains the same, while the polynomial $\PKL(x)$ simplifies to
\begin{align}
\PKL(x) &= A_0\, x^2 + c_1\, x - \frac{4 k^2}{\beta^2} A_0.
\end{align}
The range of the coordinate $x$ is given by its two real roots, ${x\in [x_-, x_+]}$. Since ${\PKL/ \Gamma}$ is the square of  the norm of the axial Killing vector $\pd_\Phi$, it has to be positive. This necessarily implies ${A_0<0}$. The roots of $\PKL(x)$ are
\begin{align} \label{eq:RootsOfPForLambdaZero}
x_{\pm} = \frac{1}{2A_0} \left( -c_1 \mp \sqrt{c_1^2 + 16 A_0^2 k^2 \beta^{-2}} \right).
\end{align}
The area of the horizon section is
\begin{align} \label{eq:TheHorizonSection}
A = \int^{\Phi_2}_{\Phi_1} \dd \Phi \int^{x_+}_{x_-}\!\!\! \sqrt{\det (g|_\mathcal{K})}\,\dd x  = (\Phi_2 - \Phi_1)(x_+ - x_-) \equiv \Delta\Phi \,(x_+ - x_-).
\end{align}

Let us consider a linear transformation between the canonical coordinates $(\zeta, \phi)$ of~\eqref{eq:CanonicalMetric} and the coordinates $(x,\Phi)$, namely
\begin{align} \label{eq:TransformationToCanonicalCoordinates}
\zeta = \omega\, x + \chi,\qquad\qquad \Phi = \lambda\, \phi + \kappa,
\end{align}
where $\omega, \chi, \lambda, \kappa$ are (not yet determined) constants. The transformed metric reads
\begin{align} \label{eq:KLmetric}
g|_\mathcal{K} = \frac{A}{4\pi}\left(\frac{4\pi\Gamma(x)}{A \omega^2 \PKL(x)}\, \dd \zeta^2 + \frac{4\pi \PKL(x)}{A \lambda^2 \Gamma(x)}\,\dd \phi^2\right),
\end{align}
where ${A=4\pi R^2}$. To ensure the same form of the metric \eqref{eq:CanonicalMetric} in the canonical coordinates, for which ${g_{\zeta\zeta}\,g_{\phi\phi}=R^4}$, the parameter $\lambda$ has to be chosen uniquely as
\begin{align}\label{eq:smallLambda}
\lambda = \frac{4\pi}{A \omega}.
\end{align}
Other constants might be found from the known range of the coordinate $\zeta$. The poles are located at $\zeta =\pm 1$ which correspond to $x = x_\pm$, hence
\begin{align}\label{eq:x+x-}
1 = \omega\, x_+ + \chi,\qquad\qquad  -1 = \omega\, x_- + \chi.
\end{align}
By using \eqref{eq:RootsOfPForLambdaZero} and \eqref{eq:smallLambda} we arrive at
\begin{align} \label{eq:omega-chi}
\omega = - \frac{2 A_0}{\sqrt{c_1^2 + 16 A_0^2 k^2\beta^{-2}}},
\qquad\qquad
\chi   = - \frac{c_1}{\sqrt{c_1^2 + 16 A_0^2 k^2\beta^{-2}}}.
\end{align}

Now we can also determine the range of the coordinate $\Phi$. The transformation \eqref{eq:TransformationToCanonicalCoordinates} gives ${\Delta \Phi = \lambda\, \Delta \phi = 2\pi \lambda}$. Using \eqref{eq:smallLambda}, \eqref{eq:TheHorizonSection} and \eqref{eq:x+x-}  we obtain ${\Delta\Phi^2 = 4\pi^2}$. Assuming naturally ${\Phi_2>\Phi_1}$ we find that ${\Delta\Phi = 2\pi}$.

Using \eqref{eq:RootsOfPForLambdaZero}, the black hole area \eqref{eq:TheHorizonSection} is thus cast into the form
\begin{align}
A = - 2\pi\, \frac{\sqrt{c_1^2 + 16 A_0^2 k^2 \beta^{-2}}}{A_0}.
\end{align}
Therefore, the coefficients \eqref{eq:omega-chi} of the transformation \eqref{eq:TransformationToCanonicalCoordinates} have a simple form in terms of the area $A$, namely
\begin{align} \label{eq:OmegaAndChi}
\omega = \frac{4\pi}{A},
\qquad\qquad
\chi   = \frac{2\pi c_1}{A A_0}.
\end{align}

The metric function ${f_\mathrm{KL}}$ can now be extracted from \eqref{eq:KLmetric} as
\begin{align} \label{eq:MetricFunctionfKL}
f_\mathrm{KL}(x) \equiv \frac{4\pi \PKL(x)}{A \lambda^2 \Gamma(x)}.
\end{align}
When we substitute all the necessary relations we get the following formula in the canonical coordinates
\begin{align} \label{eq:MetricFunctionKLZeroLambda}
f_\mathrm{KL}(\zeta) = \frac{8A_0^2\, (1-\zeta^2)}{\sqrt{c_1^2\beta^{2} + 16 A_0^2 k^2}\,(1 + \zeta^2) + 2c_1 \beta\, \zeta}.
\end{align}
The deficit angles can be now calculated using our regularity condition \eqref{eq:RegularityConditions}, yielding
\begin{align}
\delta_\pm = \frac{\pi}{2k^2} \left( \sqrt{c_1^2 \beta^2 + 16 A_0^2 k^2} \mp c_1 \beta\right) -2\pi.
\end{align}
Putting this expressions in our result \eqref{eq:MetricFunctionLambda=0} and setting ${q=0}$ we obtain exactly the function \eqref{eq:MetricFunctionKLZeroLambda}. Our result is thus fully compatible with the previous results \cite{KunduriLucietti} in this subcase.

Notice that for ${c_1 = 0}$ we obtain simply
\begin{align}\label{eq:delat+=delta-Lambda=0}
\delta_+ = \delta_- = - 2\pi \left( \frac{A_0}{|k|} + 1\right).
\end{align}
In this case we can achieve a regular geometry (${\delta_+ = 0 = \delta_-}$) by an appropriate redefinition of the range of the coordinate $\Phi$, or by a suitable choice of the ratio ${A_0/ |k|}$, which is admissible due to a freedom in the choice of one of the metric parameters.

\subsubsection{The case ${\Lambda \neq 0}$}

When the cosmological constant $\Lambda$ is non-zero the polynomial $\PKL(x)$ given by \eqref{eq:PKL(x)} is of the fourth order which considerably complicates the analytic investigation. Explicit identification of the roots $x_{\pm}$ corresponding to the poles ${\zeta_{\pm}=\pm 1}$ of the horizons with the deficit angles $\delta_{\pm}$ is not obvious, as well as the physical interpretation of the integration constants in \eqref{eq:PKL(x)} and the range of the coordinates employed in~\cite{KunduriLucietti}.

Interestingly, it is possible to complete this task in the case of uncharged extremal black holes with ${c_1=0}$. In such a case the key expression \eqref{eq:PKL(x)} becomes \emph{biquadratic}, so that it is possible to find its four roots as
\begin{align} \label{eq:RootsOfPForLambda}
x_{1,2}^2 = \frac{6}{\Lambda\beta^2}\, \Big[ A_0\beta -2\Lambda k^2 \mp \sqrt{A_0^2 \beta^2 + \tfrac{16}{3}\Lambda k^2(\Lambda k^2 - A_0\beta)}\, \Big].
\end{align}
The poles are then located at ${x_\pm = \pm x_1}$ or ${x_\pm = \pm x_2}$, depending on the precise values of the parameters and the sign of $\Lambda$. However, our further analysis is not affected by the specific choice, so let us take ${x\in [x_-, x_+] \equiv [-x_1, x_1]}$.

Now we proceed in exactly the same way as in the previous case ${\Lambda = 0}$. We assume the transformation \eqref{eq:TransformationToCanonicalCoordinates}, which results in the relations \eqref{eq:smallLambda} and \eqref{eq:OmegaAndChi}, namely
\begin{align}
\omega = \frac{2}{x_+ - x_-} = \frac{4\pi}{A} \equiv \frac{1}{R^2},
\qquad\qquad
\chi   = -\frac{x_+ + x_-}{x_+ - x_-} = 0.
\end{align}
When we put these relations into \eqref{eq:smallLambda}, we get ${\lambda = 1}$ and consequently ${\Delta \Phi = 2\pi}$. The metric function \eqref{eq:MetricFunctionfKL} now reads
\begin{align} \label{eq:MetricFunctionKLLambda}
f_\mathrm{KL}(\zeta) = \frac{\tfrac{1}{3}\Lambda \,\xi^4 - 4(A_0 \beta - 2\Lambda k^2)\, \xi^2  + 16k^2 (A_0 \beta - \Lambda k^2)}{R^2 \beta^2 (\xi^2 + 4k^2)},
\end{align}
where we have denoted ${\xi \equiv \beta  R^2} \zeta$ for brevity. The deficit angles can be calculated directly from $f_\mathrm{KL}$ or, in general, using the chain rule for the derivative of $f_\mathrm{KL}$,
\begin{align}
\frac{\dd f_\mathrm{KL}}{\dd \zeta} = \frac{1}{\omega} \frac{\dd f_\mathrm{KL}}{\dd x} = \frac{\dd}{\dd x} \frac{\PKL(x)}{\Gamma(x)} = \frac{\PKL'(x)}{\Gamma(x)} -\PKL(x) \frac{\Gamma'(x)}{\Gamma^2(x)}.
\end{align}
Using the definition \eqref{eq:RegularityConditions}, the deficit angles are thus
\begin{align}
\delta_\pm = \mp \pi\, \frac{\PKL'(x_\pm)}{\Gamma(x_\pm)} - 2\pi.
\end{align}
After substituting the functions $\Gamma(x)$ and $\PKL(x)$ from \eqref{eq:Gamma(x)}, \eqref{eq:PKL(x)}, we arrive at
\begin{align} \label{eq:DeficitAnglesKLLambda}
\delta_+ =\delta_- = 2\pi\, \frac{4k^2(2\Lambda x_+ - 1) + \beta^2 x_+^2(\tfrac{2}{3}\Lambda x_+ - 1) - 4A_0\beta x_+}{4k^2 + \beta^2 x_+^2}.
\end{align}
In this special case when ${c_1 = 0}$, \emph{the deficit angles are equal} and the black hole is non-accelerating. In fact, these conditions are \emph{equivalent}. The metric can be regularized (by a suitable redefinition of the range of the coordinate $\Phi$, or by a special choice of one of the parameters) if and only if ${c_1=0}$. For ${\Lambda=0}$, the expression \eqref{eq:DeficitAnglesKLLambda} reduces to \eqref{eq:delat+=delta-Lambda=0}.

When we put the relations \eqref{eq:DeficitAnglesKLLambda} into our main result \eqref{eq:MetricFunctionLambda}, \eqref{eq:dici} together with ${q^2=0}$, we recover \eqref{eq:MetricFunctionKLLambda}. Therefore, we have proved that $f_\mathrm{KL}$ coincides with $f_{\text{EIH}}$ for ${c_1=0}$ and the particular choice of parameters $\delta_\pm, R^2, q^2$ given by the above formulas. We have also determined the relation of the parameters of the metric \eqref{eq:KL-LineElement} to the geometric parameters of EIHs.

Considerable complications to identify the parameters of \eqref{eq:PKL(x)} in the most general case ${c_1 \ne 0}$ of extremal isolated horizons with $\Lambda \neq 0$ shows that our new form of the metric function $f_{\text{EIH}}(\zeta)$ given by \eqref{eq:MetricFunctionLambda}, whose numerator is factorized into a product of two quadratic terms, is more convenient. Moreover, because it directly contains geometrical and physical parameters, namely the deficit angles $\delta_\pm$ at the two poles of the horizon, its area $A$, and the dimensionless electromagnetic charge parameter $q^2$.
\newpage

\section{Exact type~D black holes}
\label{sec:exact type D}

A complete class of black hole spacetimes of algebraic type D with any value of the cosmological constant~$\Lambda$ and electromagnetic field (which is not null and is double aligned with the gravitation field) was presented by  Pleba\'{n}ski and Demia\'{n}ski \cite{Plebanski1976}, extending the previous work of Debever~\cite{Debever1971}. Here we employ the convenient Griffiths-Podolsk\'y form of these solutions derived in \cite{GriffithsPodolsky2006, PodolskyGriffiths2006} and summarized as Eq.~(16.18) in  \cite{Griffiths2009}, namely
\begin{align}
\dd s^2 = -\frac{1}{\Omega^2} &
  \left(-\frac{\QQ}{\rho^2}\left[\dd t- \left(a\sin^2\theta +4l\sin^2\!{\textstyle\frac{1}{2}\theta} \right)\dd\fii \right]^2 + \frac{\rho^2}{\QQ}\,\dd r^2 \right. \nonumber\\
& \quad \left. + \,\frac{\rho^2}{P}\,\dd\theta^2
  + \frac{P}{\rho^2}\,\sin^2\theta\, \big[ a\,\dd t -\big(r^2+(a+l)^2\big)\,\dd\fii \big]^2
 \right). \label{newmetricGP2005}
\end{align}
The metric functions are
\begin{align}
 \Omega &= 1 - \alpha\, \Big(\,\frac{l}{\omega}+\frac{a}{\omega}\cos\theta \Big)\, r, \nonumber\\[1mm]
 \rho^2 &= r^2 + (l + a \cos\theta)^2, \nonumber\\[2mm]
 P(\theta) &= 1-a_3\cos\theta-a_4\cos^2\theta, \nonumber\\
 \QQ (r) &= (\omega^2k+e^2+g^2)-2m\,r+\epsilon\,r^2-2\alpha\,\frac{n}{\omega}\,r^3 -\Big(\alpha^2k+\frac{\Lambda}{3}\Big)\,r^4,
  \label{nonfactorizedQ}
\end{align}
where
\begin{align}
a_3 &= 2\alpha \,\frac{a}{\omega}\,m -4\alpha^2\, \frac{a\,l}{\omega^2}\,(\omega^2k+e^2+g^2) -4\,\frac{\Lambda}{3}\,a\,l,  \nonumber\\
a_4 &= -\alpha^2\,\frac{a^2}{\omega^2}\,(\omega^2k+e^2+g^2)-\frac{\Lambda}{3}\,a^2, \label{a34}
\end{align}
while the coefficients $k$, $\epsilon$ and $n$ in \eqref{nonfactorizedQ} are determined by the relations
\begin{align}
  &\Big(\, \frac{\omega^2}{a^2-l^2}+3\alpha^2l^2 \Big)\,k
  = 1 +2\alpha\,\frac{l}{\omega}\,m -3\alpha^2\frac{l^2}{\omega^2}(e^2+g^2) -\Lambda\, l^2,
  \label{k}\\
\epsilon &= \frac{\omega^2k}{a^2-l^2} +4\alpha\,\frac{l}{\omega}\,m
 -(a^2+3l^2) \Big[\, \frac{\alpha^2}{\omega^2}\,(\omega^2k+e^2+g^2)+\frac{\Lambda}{3}\, \Big],
  \label{epsilon}\\
n &= \frac{\omega^2k\,l}{a^2-l^2} -\alpha\,\frac{a^2-l^2}{\omega}\,m
 +(a^2-l^2)\,l\, \Big[\, \frac{\alpha^2}{\omega^2}\,(\omega^2k+e^2+g^2)+\frac{\Lambda}{3} \Big].
  \label{n}
\end{align}

The metric (\ref{newmetricGP2005}) thus depends on \emph{seven usual physical parameters} $m$, $a$, $l$, $\alpha$, $e$, $g$, $\Lambda$ which characterize mass, Kerr-like rotation, NUT parameter, acceleration, electric and magnetic charges of the black hole, and the cosmological constant, respectively.

In addition, there is the twist parameter~$\omega$ related \emph{both} to $a$ and $l$ (see the discussion in \cite{Griffiths2005, GriffithsPodolsky2006}). As demonstrated in our previous works \cite{PodolskyVratny2020, PodolskyVratny2021, MatejovPodolsky}, it is very convenient to use the remaining gauge freedom to fix $\omega$ as
\begin{align}
\omega \equiv \frac{a^2 + l^2}{a}. \label{eq:ChoiceOfOmega}
\end{align}
With this choice, the general metric \eqref{newmetricGP2005} reduces directly to the familiar forms of either the Kerr--Newman-(A)dS, the Taub-NUT-(A)dS solution, or the $C$-metric with charges, rotation and the cosmological constant, without the need of further transformations, simply by setting the corresponding parameters to zero.

An important observation for our work is that \emph{horizons} are located at values of the radial coordinate $r = r_h$ which are determined by a condition
\begin{align} \label{eq:conditionForHorizonsQ}
\QQ(r_h) = 0.
\end{align}
An \emph{extremality} of the horizon is related to its degeneracy, and can be expressed as
\begin{align} \label{eq:conditionOfExtremalityQ}
\QQ'(r_h) = 0,
\end{align}
where the prime denotes the derivative with respect to $r$. As we have shown in our previous work \cite{MatejovPodolsky}, this condition is equivalent to the requirement of vanishing surface gravity ${\surg = 0}$.

The explicit form of the key metric function $\QQ (r)$ given by \eqref{nonfactorizedQ} is rather complicated when \eqref{k}--\eqref{n} are employed. It is a quartic expression in the coordinate $r$, but the coefficients are rather cumbersome. Interestingly, for ${\Lambda=0}$ it can be explicitly factorized to four roots~\cite{Griffiths2005}, thus simply identifying the corresponding horizons. This fact enabled us in \cite{MatejovPodolsky} to find and study the properties of all admitted extremal horizons.

In order to proceed with the analysis in the present case with a \emph{general cosmological constant}~$\Lambda$, we have to make an additional simplifying assumption. It turns out that we can identify the extremal horizons of \emph{all non-accelerating black holes} of algebraic type~D in the Pleba\'{n}ski and Demia\'{n}ski family.

\subsection{Non-accelerating black holes (${\alpha=0}$)}
\label{sec:nonac type D}

For vanishing acceleration, i.e.\ for the Kerr-Newman-NUT-(anti-)de~Sitter black holes, by setting ${\alpha=0}$ the expressions \eqref{a34} and \eqref{k}--\eqref{n}  with \eqref{eq:ChoiceOfOmega} considerably simplify to
\begin{align}
a_3 &= -\tfrac{4}{3}\Lambda\,a\,l,  \qquad
a_4  = -\tfrac{1}{3}\Lambda\,a^2, \nonumber\\
\frac{\omega^2 k}{a^2-l^2} &= 1 -\Lambda\,l^2,\qquad
\epsilon = 1 - \tfrac{1}{3}\Lambda\,(a^2+6l^2),\qquad
n = \big[ 1 +\tfrac{1}{3}\Lambda\,(a^2-4l^2)\big]\,l.  \label{kepsn-nona}
\end{align}
The metric \eqref{newmetricGP2005}, \eqref{nonfactorizedQ} thus reduces to
\begin{align}
\dd s^2 =
  \frac{\QQ}{\rho^2}&\left[\dd t- \left(a\sin^2\theta +4l\sin^2\!{\textstyle\frac{1}{2}\theta} \right)\dd\fii \right]^2 - \frac{\rho^2}{\QQ}\,\dd r^2  \nonumber\\
&   - \frac{\rho^2}{P}\,\dd\theta^2
  - \frac{P}{\rho^2}\,\sin^2\theta\, \big[ a\,\dd t -\big(r^2+(a+l)^2\big)\,\dd\fii \big]^2, \label{BHmetric}
\end{align}
with
\begin{align}
 \rho^2 &= r^2 + (l + a \cos\theta)^2, \nonumber\\[2mm]
 P(\theta) &= 1 + \tfrac{4}{3}\Lambda\,a\,l\,\cos\theta + \tfrac{1}{3}\Lambda\,a^2\cos^2\theta, \label{BHfunctions}\\[2mm]
 \QQ (r) &= (a^2-l^2)(1 -\Lambda\,l^2)+e^2+g^2 -2m\,r+[1 - \Lambda\,(\tfrac{1}{3}a^2+2l^2)]\,r^2 - \tfrac{1}{3}\Lambda\,r^4,
  \nonumber
\end{align}
in full agreement with Eq.~(16.23) of \cite{Griffiths2009}. Recall that this class of solutions is contained within those found in different form by Carter \cite{Carter1968b}, and that its particular subclasses were presented and discussed, e.g. by Frolov \cite{Frolov1974b} and Gibbons and Hawking \cite{GibbonsHawking1977}.

For further investigations it is useful to rewrite these black hole spacetimes in an equivalent form by introducing a coordinate
\begin{align}
\varsig = \cos\theta, \qquad \varsig\in[-1,1]. \label{eq:VarsigTheta}
\end{align}
The metric \eqref{BHmetric} then becomes
\begin{align}
\dd s^2 =
  \frac{\QQ}{\rho^2}&\left[\dd t- \left(a\,(1-\varsig^2) + 2l\,(1-\varsig)\right)\dd\fii \right]^2 - \frac{\rho^2}{\QQ}\,\dd r^2  \nonumber\\
&   - \frac{\rho^2}{\tilde{P}}\,\dd\varsigma^2
  - \frac{\tilde{P}}{\rho^2}\, \big[ a\,\dd t -\big(r^2+(a+l)^2\big)\,\dd\fii \big]^2, \label{PDmetricGP}
\end{align}
where
\begin{align}
 \rho^2 &= r^2 + (l + a\, \varsig)^2,   \label{rho-sigma}\\
 \tilde{P} (\varsig) &\equiv (1 - \varsig^2)\, P(\varsig) = (1 - \varsig^2)(1 + \tfrac{4}{3}\Lambda\,a\,l\,\varsig + \tfrac{1}{3}\Lambda\,a^2\,\varsig^2),
\label{tildeP-sigma}
\end{align}
while $\QQ (r)$ remains the same is in \eqref{BHfunctions}.

\subsection{Geometry of the horizons of non-accelerating black holes}

In our previous paper \cite{MatejovPodolsky} we investigated a class of exact spacetimes of the algebraic type~D with ${\Lambda=0}$, and we derived explicit results for a metric function which describes the geometry of extremal black hole horizons in this class. Interestingly, the derivation of these results \emph{does not differ} from the case when ${\Lambda \neq 0}$. Hence, using the formula (55) in \cite{MatejovPodolsky} (summarized in Theorem~2 of \cite{MatejovPodolsky}), the corresponding metric function reads
\begin{align}
f_{\text{D}}(\zeta) &= \frac{4\pi C^2}{A} \big[r_H^2+(a + l)^2\big]^2
    \frac{\tilde{P}(\zeta)}{\Omega^2(\zeta)\,\rho^2(\zeta)},
\label{eq:MetricFunctionTypeD}
\end{align}
where the dependence on $\Lambda$ is implicit via the specific function $\tilde{P}$. For non-accelerating black holes studied here the functions $\tilde{P}$ and $\rho$ are given by \eqref{tildeP-sigma} and \eqref{rho-sigma}, respectively, while ${\Omega=1}$ because ${\alpha=0}$. Let us recall that these functions have to be regarded as functions of a new coordinate~$\zeta$ which is related to~$\varsigma$ via
\begin{align}
\zeta(\varsig) =
\frac{\varsig - \alpha\, r_H {\displaystyle \Big(\,\frac{a}{\omega} + \frac{l}{\omega}\, \varsig \Big)}}
     {1 - \alpha\,r_H {\displaystyle \Big(\,\frac{a}{\omega}\, \varsig + \frac{l}{\omega} \Big) }}, \label{eq:ZetaAndVarsigma}
\end{align}
see Eq.~(53) in \cite{MatejovPodolsky}. However, in the present case ${\alpha=0}$ this is just an identity, ${\zeta = \varsigma}$.

The horizon area $A$ of an extremal black hole whose horizon is located at $r_H$, entering the expression \eqref{eq:MetricFunctionTypeD}, is
\begin{align} \label{eq:AreaNonAcceleratedBH}
A = 4\pi C\, [r_H^2 + (a+l)^2],
\end{align}
see Eq.~(51) in \cite{MatejovPodolsky} for the case ${\alpha=0}$.

Finally, the deficit angles around the poles are given by Eq. (57) in \cite{MatejovPodolsky},
\begin{align}
\begin{split}
\delta_+ &= 2\pi \Big(C\,(1 - a_3 - a_4) - 1 \Big),\\
\delta_- &= 2\pi \Big(C\,(1 + a_3 - a_4)\,\frac{r_H^2+(a+l)^2}{r_H^2+(a-l)^2} - 1 \Big).
\label{eq:DeficitAngles}
\end{split}
\end{align}
Recall that the free \emph{conicity parameter}~$C$ was introduced to ensure the correct range $[0,2\pi)$ of the adapted angular coordinate $\phi$.

\section{Identification of EIHs with horizons of all type D non-accelerating extremal black holes}
\label{sec:NonAcceleratingBlackHoles}

As we have already mentioned, the most important subclass of the general family of type~D black holes are solutions without acceleration (${\alpha = 0}$). In fact, these are the famous \emph{Kerr-Newman-NUT-(A)dS} black holes characterized by 6 physical parameters $m,a,l,e,g,\Lambda$. Such spacetimes in general contain \emph{two black hole horizons}, which ``merge'' when the black hole is extremal, and \emph{two cosmological horizons} due to the presence of a cosmological constant~$\Lambda$.

In view of \eqref{BHfunctions}, equation \eqref{eq:conditionForHorizonsQ} which localizes these horizons takes the form
\begin{align}\label{eq:conditionHorizonsQ}
\tfrac{1}{3}\Lambda\, r_h^4 - [1 - \Lambda\,(\tfrac{1}{3}a^2 + 2 l^2)]\,r_h^2 + 2 m\, r_h
   - (a^2 - l^2 + e^2 + g^2) + \Lambda\, l^2 (a^2 - l^2) = 0.
\end{align}
For ${\Lambda = 0}$ the condition of extremality \eqref{eq:conditionOfExtremalityQ} relates the value of the radial coordinate and the mass parameter directly as ${r_h = m}$, see \cite{MatejovPodolsky}. Inspired by this relation, we can express the mass parameter $m$ from equation \eqref{eq:conditionOfExtremalityQ} by taking the derivative of $\QQ$ given by \eqref{BHfunctions}. An algebraic manipulation leads to
\begin{align}\label{eq:m-pomoci-rH}
m = r_h - \tfrac{1}{3}\Lambda\, (a^2 + 6\,l^2 + 2r_h^2)\, r_h.
\end{align}
When we substitute this relation back into \eqref{eq:conditionHorizonsQ} we obtain
\begin{align} \label{eq:ForCoordinateOfHorizon}
 (r_h^2 + l^2)(\Lambda\,r_h^2  + \Lambda\,l^2  - 1) + \tfrac{1}{3}\Lambda\, a^2 (r_h^2 - 3l^2)
 + a^2 + e^2 + g^2  = 0.
\end{align}
Interestingly, this is a \emph{quadratic equation} for $r_h^2$ whose distinct two roots are
\begin{align}
\begin{split}
r_H^2 &= \frac{1}{2\Lambda}\Big[1 - \Lambda\,(\tfrac{1}{3}a^2+2l^2) - \sqrt{D}\,\Big], \\
r_C^2 &= \frac{1}{2\Lambda}\Big[1 - \Lambda\,(\tfrac{1}{3}a^2+2l^2) + \sqrt{D}\,\Big],
\label{eq:r_andr_c}
\end{split}
\end{align}
where
\begin{align} \label{eq:DedD}
D  \equiv 1 - \Lambda\,(\tfrac{14}{3}a^2+4e^2+4g^2) + \Lambda^2 a^2\, (\tfrac{1}{9} a^2+\tfrac{16}{3} l^2).
\end{align}
The first root $r_H$ represents a \emph{black hole} horizon, while $r_C$ localizes a \emph{cosmological} horizon. To see this directly, let us compute the area of the two surfaces. Substituting these values of $r_H$ and $r_C$ into \eqref{eq:AreaNonAcceleratedBH} gives
\begin{align}
\begin{split}
A_H &=  \frac{2\pi C}{\Lambda}\Big[1 + \Lambda\,a\,(\tfrac{5}{3} a + 4 l) - \sqrt{D}\,\Big], \\
A_C &=  \frac{2\pi C}{\Lambda}\Big[ 1 + \Lambda\,a\,(\tfrac{5}{3} a + 4 l) + \sqrt{D}\,\Big],
\label{eq:A_HandA_c}
\end{split}
\end{align}
respectively. Expansion for small values of $\Lambda$ leads to
\begin{align}
\begin{split}
A_H &= 4\pi C\, \big[2a(a + l) + e^2 + g^2\big] + \mathcal{O}(\Lambda), \\
A_C &= 4\pi C\, \frac{1}{\Lambda} + \mathcal{O}(1).
\end{split}
\end{align}
In the limit of asymptotically flat spacetime ${\Lambda \rightarrow 0}$, the area $A_C$ diverges, i.e.\ the cosmological horizon expands to infinity. On the other hand, in this limit the black hole horizon has the area ${A_H = 4\pi C\, \big[ a^2 - l^2 + e^2 + g^2 + (a + l)^2 \big] = 4\pi C\, \big[ r_H^2 + (a + l)^2 \big]}$ and ${m=r_H}$, which fully agrees with Eqs.~(110) and~(109) of \cite{MatejovPodolsky}, respectively.

From \eqref{eq:r_andr_c} it is obvious that for each ${r_H^2}$ and ${r_C^2}$ there actually exists a \emph{pair of horizons}, namely ${\pm r_H}$ and ${\pm r_C}$. There are thus extremal horizons in both regions ${r>0}$ and ${r<0}$. Moreover, it can be seen from \eqref{eq:m-pomoci-rH} that ${r_h \to - r_h}$ corresponds to ${m \to -m}$. By substituting ${\pm r_H}$ from \eqref{eq:r_andr_c} into \eqref{eq:m-pomoci-rH} we obtain an explicit expression ${m(a,l,e,g,\Lambda)}$ determining the value of the mass parameter for the corresponding extremal black hole horizon.

\emph{The precise number and degeneracy of these extremal horizons} in the Kerr-Newmann-NUT-(A)dS spacetime depend on the cosmological constant $\Lambda$ (primarily divided into the distinct ${\Lambda<0}$ and ${\Lambda>0}$ cases) and  on specific values of the physical parameters $a, l, e, g$. In the Appendix we carefully discuss all the possibilities. Let us summarize here only the main results:

\begin{itemize}

\item In the ${\Lambda<0}$ case there is no cosmological horizon. The extremal black hole horizon is located at $r_H$ given by \eqref{eq:BHhorizonLambda<0}, provided the NUT parameter~$l$ satisfies the condition \eqref{eq:MaxNUTLambda<0}.

\item In the ${\Lambda>0}$ case the admittable values of the cosmological constant form a discontinuous interval
${\Lambda \in (0,\Lambda^-]\cup(\Lambda^+, \infty)}$, where $\Lambda^\pm$ are given by \eqref{eq:LambdaPlusMinus}.

\item The boundary value $\Lambda^-$ characterizes a situation in which all horizons merge into one multiple-degenerate horizon located at $r_H = r_C$ given by \eqref{eq:rH=rC}. Moreover, the NUT parameter~$l$ has to fulfil the condition \eqref{eq:UpperBoundForl}.

\item For ${\Lambda \in (0,\Lambda^-)}$ there is the extremal black hole horizon as well as the cosmological horizon at $r_H$ and $r_C$ expressed by \eqref{eq:rHAndrC} and \eqref{eq:rHAndrD}, respectively. The value of $l$ is again restricted by \eqref{eq:UpperBoundForl}. Depending on the relative values of $|a|$ and $|l|$, the cosmological constant~$\Lambda$ is further restricted by \eqref{eq:LowerLambda}, or is not restricted at all.

\item On the other hand, existence of the extremal black hole horizon is automatically excluded when ${\Lambda \in (\Lambda^+, \infty)}$. In this case, the cosmological horizon is present only if ${|l|<|a|}$ and $\Lambda$ is greater than $\Lambda_0$ given by \eqref{eq:Lambda0}.

\end{itemize}

Let us now return to the main topic of this section which is the identification of the metric functions
$f_{\text{EIH}}(\zeta)$ and $f_{\text{D}}(\zeta)$ of extremal black holes. The former is given by \eqref{eq:MetricFunctionLambda} while the latter by \eqref{eq:MetricFunctionTypeD}. For non-accelerating type D black holes it simplifies to
\begin{align} \label{eq:MetricFunctionNonAccelerating}
f_{\mathrm{D}} = C\,[r_H^2 + (a + l)^2]\,(1-\zeta^2)\, \frac{1 + \tfrac{4}{3}\Lambda al\,\zeta + \tfrac{1}{3}\Lambda a^2\, \zeta^2 }{r_H^2+(l + a\zeta)^2},
\end{align}
and the deficit angles \eqref{eq:DeficitAngles} around the poles are
\begin{align}
\begin{split}
\delta_+ &= 2\pi C\, [1 + \tfrac{1}{3}\Lambda a(a+4l)] - 2\pi, \\
\delta_- &= 2\pi C\, [1 + \tfrac{1}{3}\Lambda a(a - 4l)]\, \frac{r_H^2+(a+l)^2}{r_H^2+(a-l)^2} - 2\pi.
\label{eq:DeficitAnglesNonAccelerating}
\end{split}
\end{align}
To keep the relations compact and readable, we do not substitute for $r_H$ from \eqref{eq:r_andr_c}.

To complete the investigation of the extremal isolated horizons in the full family of Kerr-Newman-NUT-(A)dS black holes, we substitute the values \eqref{eq:DeficitAnglesNonAccelerating} for $\delta_\pm$ together with the relation ${R^2=C\, [r_H^2+(a+l)^2]}$, see \eqref{eq:AreaNonAcceleratedBH}, into the formula \eqref{eq:MetricFunctionLambda}, \eqref{eq:dici} for $f_{\mathrm{EIH}}$, and we compare the resulting function with \eqref{eq:MetricFunctionNonAccelerating}. It turns out that it is possible to match $f_{\mathrm{D}}$ and $f_{\mathrm{EIH}}$ exactly by a \emph{unique choice} of the dimensionless charge parameter $q^2$, namely by
\begin{align}
q^2 = 4\pi C\, \frac{(r_H^2+l^2)\big[1-\Lambda(r_H^2+l^2)\big] - a^2 \big[1+\Lambda(\tfrac{1}{3}r_H^2-l^2)\big]}{r_H^2 + (a-l)^2}.
\label{eq:q^2generic}
\end{align}
 Indeed, for these values of $\delta_\pm$, $R^2$ and $q^2$ we obtain
\begin{align}
c_0 & = \Xi\, (r_H^2+l^2),
\hspace{10.3mm}
d_0 = C\,\Xi\,[r_H^2+(a+l)^2],  \nonumber\\[1mm]
c_1 & = \Xi\,2al,
\hspace{21mm}
d_1 = C\,\Xi\,[r_H^2+(a+l)^2]\,\tfrac{4}{3}\Lambda a l , \nonumber\\[1mm]
c_2 & = \Xi\, a^2,
\hspace{22.8mm}
d_2 = C\,\Xi\,[r_H^2+(a+l)^2]\,\tfrac{1}{3}\Lambda a^2 , \nonumber
\end{align}
where
\begin{align}
\Xi = 8\pi C\, \frac{ 1-\tfrac{1}{3}\Lambda(2r_H^2+2l^2+a^2) }{r_H^2+(a-l)^2},
\label{eq:DefXi}
\end{align}
so that \eqref{eq:MetricFunctionLambda} is exactly the function \eqref{eq:MetricFunctionNonAccelerating}.

Moreover, using the definition \eqref{eq:DEf-g^2} of $q^2$ for \eqref{eq:q^2generic} and the area ${A=4\pi R^2}$ of the extremal black hole horizon at $r_H$ given by \eqref{eq:AreaNonAcceleratedBH}, we arrive at the explicit relation between the physically defined charges \eqref{eq:PhysicalCharges} and the parameters of the type~D metric  \eqref{BHmetric} as
\begin{align}
Q_E^2 + Q_M^2 = C^2 \,\frac{r_H^2 + (a+l)^2}{r_H^2 + (a-l)^2}
\Big[(r_H^2+l^2)\big[1-\Lambda(r_H^2+l^2)\big] - a^2 \big[1+\Lambda(\tfrac{1}{3}r_H^2-l^2)\big]\Big].
\end{align}
Expressing $r_H$ using \eqref{eq:r_andr_c}, that is ${r_H^2 + l^2 = \big(1 - \tfrac{1}{3}\,\Lambda\,a^2 - \sqrt{D}\,\big)/(2\Lambda)}$, we get
\begin{align}
Q_E^2 + Q_M^2 = C^2 \,
  \frac{1 + \Lambda\,a\,(\tfrac{5}{3}a + 4 l) - \sqrt{D}}
       {1 + \Lambda\,a\,(\tfrac{5}{3}a - 4 l) - \sqrt{D}}
       \,(e^2 + g^2),
\label{eq:QeQm-eg}
\end{align}
where $D$ is given by  \eqref{eq:DedD}.
The physical charges $Q_E, Q_M$ are thus \emph{directly related} to the metric charge parameters $e, g$, although \emph{they are not identical}. However, a simple relation ${Q_E^2 + Q_M^2 = C^2 (e^2 + g^2)}$ is recovered if (and only if) ${a\,l\,\Lambda=0}$, i.e.\ when the Kerr rotation vanishes (${a=0}$), when the NUT parameter vanishes (${l=0}$), or in the absence of the cosmological constant (${\Lambda=0}$).

We can thus summarize the results in the following theorem.

\begin{theorem} \label{th:Equivalence}
Extremal horizons in the complete family of Kerr-Newman-NUT-(A)dS black holes (all extremal black holes of algebraic type D without acceleration) are located at $r_H$ determined by \eqref{eq:r_andr_c}. Their geometry is represented by the induced metric of the form \eqref{eq:CanonicalMetric}, where the metric function $f_{\mathrm{D}}$ is given by \eqref{eq:MetricFunctionNonAccelerating}.

Moreover, this function precisely coincides with the metric function $f_{\text{EIH}}(\zeta)$ of axisymmetric extremal isolated horizons (EIHs) in asymptotically (A)dS spacetime, given in
 Theorem~\ref{th:MetricIHwithLambda}. The  geometric parameters of EIHs are identified with the parameters of the metric \eqref{BHmetric} via the relation ${R^2=C\, [r_H^2+(a+l)^2]}$, the deficit angles $\delta_+$, $\delta_-$ around the poles are given by \eqref{eq:DeficitAnglesNonAccelerating}, and the physical charges $Q_E$, $Q_M$ are given by \eqref{eq:QeQm-eg}.
\end{theorem}

\subsection{Kerr-Newman-(A)dS black holes (${l=0}$)}
\label{sec:KNAdS}

Let us have a closer look at the physically most relevant case, when the black hole represents a charged and rotating mass in (anti-)de Sitter spacetime without the NUT parameter.

The black hole extremal horizon ${r_H>0}$ (and ${r_H<0}$) is located at the radial coordinate
\begin{align}
r_H^2 &= \frac{1}{2\Lambda}\left[1 - \tfrac{1}{3} \Lambda a^2 - \sqrt{1 - \Lambda(\tfrac{14}{3} a^2 + 4e^2 + 4g^2) + \tfrac{1}{9} \Lambda^2 a^4}\,\right],
\end{align}
see \eqref{eq:r_andr_c}. By setting ${l=0}$ in \eqref{eq:MetricFunctionNonAccelerating}, the metric function simplifies to
\begin{align}\label{eq:fD-KerrNewmanAdS}
f_{\mathrm{D}} = C(r_H^2 + a^2)(1-\zeta^2)\, \frac{1 + \tfrac{1}{3}\Lambda\, a^2\, \zeta^2 }{r_H^2 + a^2\, \zeta^2}.
\end{align}
The deficit angles \eqref{eq:DeficitAnglesNonAccelerating} remain non-zero, namely
\begin{align}
\delta_+ =\delta_- = 2\pi C\,(1 + \tfrac{1}{3}\Lambda\, a^2) - 2\pi,
\end{align}
but for a unique choice of the conicity parameter
\begin{align}\label{eq:Cregular}
C = (1 + \tfrac{1}{3}\Lambda\, a^2)^{-1},
\end{align}
we obtain a solution with \emph{both poles regular}.

Then the function $f_{\mathrm{D}}(\zeta)$ given by \eqref{eq:fD-KerrNewmanAdS} has precisely the form of \eqref{eq:MetricFunctionRegularAxis} of $f_{\text{EIH}}(\zeta)$, with
\begin{align}
q^2 &= 4\pi\, \frac{r_H^2 - a^2 - \tfrac{1}{3}\Lambda\, r_H^2 (3r_H^2 + a^2)}{(r_H^2 + a^2)(1 + \tfrac{1}{3}\Lambda\, a^2)}\nonumber\\[2mm]
&= \frac{8\pi \Lambda\, (e^2 + g^2)}{(1 + \tfrac{1}{3}\Lambda\, a^2)\, \Big(1 + \tfrac{5}{3} \Lambda\,a^2 - \sqrt{1 - \Lambda\, (\tfrac{14}{3}a^2 + 4e^2 + 4g^2) + \tfrac{1}{9}\Lambda^2 a^4}\,\Big)}.
\end{align}
This is consistent with \eqref{eq:q^2generic}. Notice also that the limit ${\Lambda \rightarrow 0}$ is well-defined and non-zero,
\begin{align}
\lim_{\Lambda \rightarrow 0} q^2 = 4\pi \,\frac{e^2 + g^2}{2 a^2 + e^2 + g^2}.
\end{align}

Due to  \eqref{eq:DEf-g^2} and \eqref{eq:A_HandA_c}, the relation between the charge parameters is
\begin{align} \label{eq:RelationBetweenChargesl=0}
Q_E^2 + Q_M^2  = C^2 (e^2 + g^2) = \frac{e^2 + g^2}{(1 + \tfrac{1}{3}\Lambda\, a^2)^2}.
\end{align}
The genuine electric and magnetic charges $Q_E$, $Q_E$  are thus proportional to the metric charge parameters $e$, $g$. However, the proportionality factor $C$ determining the conicity is now fixed by the condition \eqref{eq:Cregular} to achieve ${\delta_+ =0=\delta_-}$, i.e. regular both axes.

\subsection{Charged NUT-(A)dS black holes (${a=0}$)}
\label{sec:NonRotatingBlackHoles}

In this part we concentrate on non-rotating black holes characterized by a condition ${a=0}$. In fact, in this case \emph{necessarily} ${\alpha = 0}$, because there is no accelerating NUT solution in the considered class of type D spacetimes \cite{Griffiths2005}.

The equation \eqref{eq:ForCoordinateOfHorizon} simplifies to
\begin{align}
 (r_H^2 + l^2)(\Lambda\,r_H^2 + \Lambda\,l^2  - 1 ) + e^2 + g^2  = 0,
\end{align}
with explicit solutions
\begin{align}
r_H^2 &= \frac{1 - \sqrt{1- 4\Lambda (e^2+g^2)}}{2\Lambda} - l^2 , \label{eq:r_Hcharget NUTAdS}  \\
r_C^2 &= \frac{1 + \sqrt{1- 4\Lambda (e^2+g^2)}}{2\Lambda} - l^2 .  \nonumber
\end{align}
Expansion for small values of the cosmological constant yields
\begin{align}
r_H^2 = e^2 + g^2 - l^2 + \mathcal{O}(\Lambda), \hspace{15mm}
r_C^2 =  \frac{1}{\Lambda} + \mathcal{O}(1),
\end{align}
so we immediately recognize the \emph{black hole} horizons at $r_H$ and the \emph{cosmological} horizons at $r_C$.

The area of the black hole horizon given by \eqref{eq:AreaNonAcceleratedBH} is
\begin{align} \label{eq:AreaNonAcceleratedNonrotBH}
A = 2\pi C\, \frac{1 - \sqrt{1- 4\Lambda (e^2+g^2)}}{\Lambda},
\end{align}
which for ${\Lambda \to 0}$ reduces to ${4\pi C\,(e^2+g^2)}$, in agreement with the results of \cite{MatejovPodolsky}.
Notice also that in absence of electric and magnetic charge ${e=0=g}$, the black hole horizon can not be extremal.

Under the current assumption of a non-rotating black hole, the metric function \eqref{eq:MetricFunctionNonAccelerating} simplifies considerably to
\begin{align} \label{eq:MetricFunctionRegular}
f_{\text{D}}(\zeta) = C\, (1 - \zeta^2).
\end{align}
This result does not depend on $\Lambda$, and is the same as in the case when ${\Lambda = 0}$. Geometry of the black hole horizon is that of a quasi-regular sphere. The deficit angles \eqref{eq:DeficitAngles} around the poles are zero provided ${C=1}$ because ${a_3=0=a_4}$, and thus
\begin{align}
\delta_+ =\delta_- = 2\pi (C-1).
\end{align}

In order to map the metric function $f_{\text{EIH}}(\zeta)$ to $f_{\text{D}}(\zeta)$, we substitute the above relations into \eqref{eq:dici} which gives
\begin{align}
d_0 & = 8\pi C^2
     + \tfrac{1}{3}\Lambda R^2 \big[ 4\pi (\Lambda R^2-5C) + q^2 \big], & c_0 & = 4\pi ( C-\tfrac{1}{3}\Lambda R^2) + q^2, \nonumber\\[1mm]
d_1 & = 0, & c_1 & = 0,  \\[1mm]
d_2 & = \tfrac{1}{3}\Lambda R^2 \big[ 4\pi (C-\Lambda R^2) - q^2 \big], & c_2 & = 4\pi (C-\Lambda R^2) - q^2,  \nonumber
\end{align}
so that the metric function \eqref{eq:MetricFunctionLambda} becomes
\begin{align}
  f_{\text{EIH}}(\zeta) &= (1-\zeta^2)\,  \frac{ 2 C^2 + \tfrac{1}{3}\Lambda R^2 \big[ (\Lambda R^2-5C + \tfrac{1}{4\pi } q^2)
   + (C - \Lambda R^2 - \tfrac{1}{4\pi } q^2)\,\zeta^2  \big]}
  { (C-\tfrac{1}{3}\Lambda R^2 + \tfrac{1}{4\pi } q^2) + (C-\Lambda R^2 - \tfrac{1}{4\pi } q^2 )\,\zeta^2 }.
\end{align}
It reduces to \eqref{eq:MetricFunctionRegular} if (and only if) we choose
\begin{align} \label{eq:qFora=0}
q^2 = 4\pi(C - \Lambda R^2) = 4\pi C \,[1 -  \Lambda\,(r_H^2+l^2)].
\end{align}
Using the definition \eqref{eq:DEf-g^2} of $q^2$ and substituting for $r_H$ from \eqref{eq:r_Hcharget NUTAdS}  and \eqref{eq:AreaNonAcceleratedBH} we arrive at
\begin{align}
Q_E^2 + Q_M^2  = C^2 (e^2 + g^2).
\end{align}
The physically defined charges $Q_E$, $Q_M$ are thus directly proportional to the electric and magnetic parameters $e$, $g$ of the type D metric via the conicity $C$.

\section{Conclusion}

The main aim of this paper was to extend the results from our previous work \cite{MatejovPodolsky} in which we investigated in detail the unique properties of axially symmetric \emph{extremal} isolated horizons (EIHs) in asymptotically flat spacetimes. Here we considered such horizons in asymptotically (anti-)de Sitter spacetimes with non-zero cosmological constant ${\Lambda \neq 0}$.

After we introduced in Sec.~2 the necessary notation and basic definitions we systematically studied constrain equations following from the NP formalism.
We concluded that the electromagnetic field, represented by tetrad projections $\phi_{i}$, and the spin coefficient $\pinp$ remain unchanged compared to the case with ${\Lambda = 0}$. Namely, in the natural coordinates $\zeta$ and $\phi$  adapted to the horizon geometry they are given explicitly as
\begin{align}
\pinp \eqNH \sqrt{\frac{f}{2}} \,\frac{1}{R\,(\zeta + c_{\pi})},
\qquad
\phi_1 \eqNH \frac{c_{\phi}}{(\zeta + c_{\pi})^2},
\end{align}
see equation \eqref{eq:PhiAndPiNP}.
Using these results, we were able to integrate the remaining equation \eqref{eq:EquationForF} constraining the horizon geometry. Our first main result of this paper is summarized in Theorem~\ref{th:MetricIHwithLambda}. In particular, the metric function $f_{\text{EIH}}$ describing the induced metric on the horizon reads
\begin{align}
  f_{\text{EIH}}(\zeta) &= (1-\zeta^2)\,  \frac{ d_0 + d_1 \,\zeta + d_2\,\zeta^2 }{ c_0 + c_1 \,\zeta + c_2\,\zeta^2 },
\end{align}
where the constants $d_i, c_i$ are given in \eqref{eq:dici}. The function is unique, well-behaved, and depends on \emph{6 real independent parameters}, namely two deficit angles $\delta_+, \delta_-$ at the horizon poles, the square of the radius  $R^2$ (the horizon area $A$ divided by $4\pi$),  the cosmological constant $\Lambda$, and the total electric and magnetic charges $Q_E, Q_M$. It further simplifies for various special choices of these parameters. For instance, we recover the recently derived solution \eqref{eq:BLMetricFunction} by Buk and Lewandowski \cite{BukLewandowski} when the function $f_{\text{EIH}}$ is assumed to be regular at both poles (${\delta_+ = 0 = \delta_-}$). For ${\Lambda = 0}$ it precisely reduces to the solution \eqref{eq:MetricFunctionLambda=0} which we investigated in \cite{MatejovPodolsky}.

We also compared our result \eqref{eq:MetricFunctionLambda} with an analogous, previously known result \eqref{eq:KL-inducedLineElement}, which was derived in the context of near horizon geometries \cite{KunduriLucietti, KunduriLucietti2009a, LiLucietti2013, KunduriLucietti2009b}. In two special cases of uncharged black holes (when ${\Lambda = 0}$ and ${c_1 = 0}$, respectively) we proved the equivalence of the results. Furthermore, we discussed advantages of our approach which leads to a more elegant form with integration constants having a direct geometrical interpretation and with the full gauge freedom already fixed.

Our second objective here was to compare the general result \eqref{eq:MetricFunctionLambda} with the horizon geometry of extremal black holes in the Pleba\'{n}ski and Demia\'{n}ski class of exact solutions of the algebraic type~D. It is represented by the line element \eqref{newmetricGP2005} in a convenient parametrization by Griffiths and Podolsk\'{y} \cite{GriffithsPodolsky2006, PodolskyGriffiths2006}. In \cite{MatejovPodolsky} we derived a specific metric function $f_{\text{D}}$, which describes the geometry of the horizon of such type~D black holes. Its general form \eqref{eq:MetricFunctionTypeD} is not affected by any value of $\Lambda$, since it enters the function only indirectly via $\tilde{P}, A, r_H$. Hence the formula \eqref{eq:MetricFunctionTypeD} remains valid also for ${\Lambda \neq 0}$.

For reasons of simplicity, we restricted our subsequent analysis to \emph{non-accelerating} black holes with ${\alpha = 0}$, that is to the family of Kerr-Newman-NUT-(anti-)de~Sitter black holes \eqref{BHmetric}, \eqref{BHfunctions}.

We identified two types of extremal horizons --- the \emph{black hole} one and a the \emph{cosmological} one. They are located at radial coordinates $\pm r_H$ and $\pm r_C$, respectively, expressed by \eqref{eq:r_andr_c}. Their precise number and degeneracy depend on the cosmological constant $\Lambda$ (primarily divided into the distinct cases ${\Lambda<0}$ and ${\Lambda>0}$), as it is carefully analysed in the Appendix.

In the last part of Sec.~\ref{sec:NonAcceleratingBlackHoles} of our work we were able to show that the function $f_{\text{D}}$ has the \emph{same form} as $f_{\text{EIH}}$ for every combination of the physical parameters. The result is summarized in Theorem \ref{th:Equivalence}.
The metric function $f_{\text{D}}$ is simplified to \eqref{eq:MetricFunctionNonAccelerating}, namely
\begin{align}
f_{\mathrm{D}} = C\,[r_H^2 + (a + l)^2]\,(1-\zeta^2)\, \frac{1 + \tfrac{4}{3}\Lambda al\,\zeta + \tfrac{1}{3}\Lambda a^2\, \zeta^2 }{r_H^2+(l + a\zeta)^2}.
\end{align}
This function is equivalent to $f_{\mathrm{EIH}}$ if we choose the dimensionless charge parameter $q^2$ as
\begin{align}
q^2 = 4\pi C\, \frac{(r_H^2+l^2)\big[1-\Lambda(r_H^2+l^2)\big] - a^2 \big[1+\Lambda(\tfrac{1}{3}r_H^2-l^2)\big]}{r_H^2 + (a-l)^2},
\end{align}
see \eqref{eq:q^2generic}. The key observation is that it does not depend on the coordinate $\zeta$, thus it can be regarded as a different parametrization of the \emph{same} function. The reason why we had to find its (unique) value is that the parameters of the metric \eqref{BHmetric} have well-understood meanings only in special cases. Applying the definition \eqref{eq:DEf-g^2}, we thus obtained a non-trivial relation between the genuine electric and magnetic charges \eqref{eq:PhysicalCharges} and the charge parameters of the Kerr-Newman-NUT-(anti-)de~Sitter metric, namely
\begin{align}
Q_E^2 + Q_M^2 = C^2 \,
  \frac{1 + \Lambda\,a\,(\tfrac{5}{3}a + 4 l) - \sqrt{D}}
       {1 + \Lambda\,a\,(\tfrac{5}{3}a - 4 l) - \sqrt{D}}
       \,(e^2 + g^2),
\end{align}
where $D$ is given by  \eqref{eq:DedD}. The charges are mutually proportional, ${Q_E^2 + Q_M^2 = C^2 (e^2 + g^2)}$, if and only if ${a\,l\,\Lambda=0}$.

In Sec.~\ref{sec:KNAdS} we concentrated on the physically most relevant subcase when ${l=0}$. We found that the poles are \emph{not generally regular}, although they can be regularized by a suitable choice \eqref{eq:Cregular} of the conicity parameter $C$. Due to this choice there is a specific relation \eqref{eq:RelationBetweenChargesl=0} between the electric and magnetic charges, which simplifies to equality ${Q_E^2 + Q_M^2 = e^2 + g^2}$ in asymptotically flat spacetimes.

Another interesting example was discussed in Sec.~\ref{sec:NonRotatingBlackHoles}. It represents the most general non-rotating (${a=0}$) charged NUT black hole of type~D in the (anti-)de~Sitter background. The intrinsic geometry of its horizon is identical to the geometry of a quasi-regular sphere \eqref{eq:MetricFunctionRegular}, and it does not depend on any parameter apart from the free conicity parameter~$C$. Though not obvious, the function $f_{\mathrm{EIH}}$ also admits this possibility, and it appears when the dimensionless charge parameter $q^2$ has the particular value given by \eqref{eq:qFora=0}.


\section*{Acknowledgements}

This paper was supported by the Czech Science Foundation Grant No.~GA\v{C}R 20-05421S.

\newpage

\section*{Appendix: Analysis of the number and degeneracy of the extremal horizons of non-accelerating black holes}

To simplify the analysis, let us denote
\begin{align}\label{eq:x=rh2}
x \equiv r_h^2,
\end{align}
and rewrite the key equation \eqref{eq:ForCoordinateOfHorizon} for the position of the horizons in the standard form
\begin{align} \label{eq:QuadraticEquationForX}
p\, x^2 + q\, x + s = 0,
\end{align}
where the constants are
\begin{align}\label{eq:pqs}
p &= \Lambda, \nonumber\\
q &= - 1 + \Lambda\, ( \tfrac{1}{3}a^2 + 2l^2), \\
s &=a^2 + e^2 + g^2 - l^2 + \Lambda\, l^2 (l^2 - a^2). \nonumber
\end{align}
The solution of this quadratic equation is ${x_{\pm} = (-q \pm \sqrt{D})/(2p)}$, that is
\begin{align}\label{eq:x_pmAppendix}
x_{\pm} = \frac{ 1}{2\Lambda}\,\Big[1 - \Lambda\, ( \tfrac{1}{3}a^2 + 2l^2) \pm \sqrt{D}\,\Big],
\end{align}
where the discriminant ${D \equiv q^2 - 4ps}$ reads
\begin{align}\label{eq:Discriminant}
D = 1 - \Lambda\, (\tfrac{14}{3}a^2 + 4 e^2 + 4g^2) + \Lambda^2a^2 (\tfrac{1}{9}a^2 + \tfrac{16}{3} l^2).
\end{align}
It is a quadratic expression in $\Lambda$.

In order to have a well-defined coordinate position of a horizon $r_h$ by \eqref{eq:x=rh2}, the corresponding root has to be non-negative, ${x\geq 0}$.

The special case when ${x = 0 = r_h}$ implies ${s=0}$, which appears whenever the cosmological constant takes the special value
\begin{align}\label{eq:Lambda0}
\Lambda_0 = \frac{ a^2  - l^2  + e^2 + g^2}{l^2\, (a^2 - l^2)}.
\end{align}
In the uncharged case, ${\Lambda_0=l^{-2}}$.

The number of real roots $x$ is determined by the sign of~$D$ in \eqref{eq:x_pmAppendix}. This discriminant vanishes for certain values of $\Lambda$, namely
\begin{align} \label{eq:LambdaPlusMinus}
\Lambda^\pm = 3\, \frac{7a^2+6e^2+6g^2 \pm \sqrt{(7a^2+6e^2+6g^2)^2 - a^2(a^2+48l^2)}}
{a^2(a^2+48l^2)}.
\end{align}
Such values ${\Lambda^\pm}$ are real and positive provided
$a^2(a^2+48l^2)<(7a^2+6e^2+6g^2)^2$, i.e.
\begin{align}\label{eq:Lambda-pm-real>0}
l^2 < a^2 + \frac{7}{4} (e^2+g^2) + \frac{3}{4}\,\frac{(e^2+g^2)^2}{a^2}.
\end{align}
For uncharged black holes this condition is simply ${l^2<a^2}$.

For \emph{negative} values of $\Lambda$ the discriminant \eqref{eq:Discriminant} is always positive, while for \emph{positive}~$\Lambda$ it acquires negative, positive and zero values. The case ${\Lambda = 0}$ was investigated in our previous work \cite{MatejovPodolsky}. Thus, we restrict our attention to the remaining cases ${\Lambda \lessgtr 0}$, which we will discuss separately.

\subsection*{The case ${\Lambda <0}$}

Since ${D>0}$, there are always \emph{two real roots} $x_{\pm}$ given by \eqref{eq:x_pmAppendix}. Due to \eqref{eq:x=rh2} these have to be positive. It is easy to infer that ${x_+ < 0}$ for any combination of the metric parameters (indeed, ${p<0}$ and ${-q >0}$, so that ${x_+ >0}$ implies ${-q + \sqrt{D}<0}$ which is a contradiction). On the other hand, from ${x_- >0}$ we obtain a non-trivial constraint
\begin{align*}
-q < \sqrt{D} = \sqrt{q^2 - 4ps} \quad \Leftrightarrow \quad s > 0 \quad \Leftrightarrow \quad
\Lambda\, l^4 - (1 + \Lambda a^2)\,l^2 + a^2 + e^2 + g^2 > 0.
\end{align*}
If ${l=0}$, the last inequality holds for any $a, e, g$. Hence, we may regard it as a restriction imposed on the admittable values of $l$. It is a quadratic polynomial in $l^2$ with two roots
\begin{align}
(l^2)_\pm = \frac{1+\Lambda a^2 \pm \sqrt{(1+\Lambda a^2)^2 + 4(-\Lambda)(a^2 + e^2 + g^2)}}{2\Lambda}.
\end{align}
Now, ${(l^2)_+ < 0}$ (since for ${\Lambda<0}$ the expression under the square root is a sum of positive numbers) which is forbidden. The second root $(l^2)_-$ defines a maximal  range of~$l$
\begin{align} \label{eq:MaxNUTLambda<0}
l \in (- l_{\text{max}}, l_{\text{max}}),  \qquad l_{\text{max}} \equiv \sqrt{(l^2)_-}.
\end{align}
In the \emph{uncharged case} when ${e=0=g}$ we obtain ${(l^2)_-=a^2}$, so that the interval is simply ${|l|<|a|}$.

To summarize, in the case ${\Lambda<0}$ there is \emph{no cosmological horizon} (which would be at~${r_C^2 \equiv x_+}$) while the \emph{extremal black hole horizon} is located at
\begin{align} \label{eq:BHhorizonLambda<0}
r_H^2 \equiv x_-&= \frac{1}{2\Lambda}\Big[1 - \Lambda\,(\tfrac{1}{3}a^2+2l^2) - \sqrt{D}\,\Big],
\end{align}
see \eqref{eq:x_pmAppendix}, where the discriminant is given by \eqref{eq:Discriminant}.

For the special value  ${\Lambda=\Lambda_0<0}$ of the cosmological constant given by \eqref{eq:Lambda0} with ${a^2 < l^2 < a^2+e^2+g^2}$,  we obtain ${r_H=0}$. This also admits the non-rotating case ${a=0}$.

\subsection*{The case ${\Lambda > 0}$ with ${\Lambda = \Lambda^\pm}$}

When ${D = 0}$, \emph{all horizons} merge into \emph{one multiple-degenerate horizon}. The corresponding solution for ${x_0 \equiv x_+ = x_-}$ is
\begin{align}
x_0^\pm = \frac{1}{2\Lambda^\pm} - \tfrac{1}{6}a^2-l^2.
\end{align}
Positivity of this root requires ${\,\Lambda^\pm (\tfrac{1}{3}a^2 + 2 l^2)<1}$.

This is \emph{violated by}~$\Lambda^+$, as demonstrated by the following estimate:
\begin{align} \label{eq:EstimateForLamda}
& \Lambda^+ (\tfrac{1}{3}a^2 + 2 l^2)  \nonumber \\
& = \frac{7a^2+6e^2+6g^2 + \sqrt{(7a^2+6e^2+6g^2)^2 - a^2(a^2+48l^2)}}
{a^2(a^2+48l^2)}\, (a^2 + 6l^2)  \nonumber \\
& \geq \frac{7a^2 + \sqrt{48 a^4 - 48a^2l^2}}{a^2(a^2+48l^2)}\, (a^2 + 6l^2) =  F(\xi),
\end{align}
where we have introduced
\begin{align} \label{eq:DefF}
\xi \equiv \frac{l^2}{a^2} \ge 0,\qquad \qquad
F(\xi) \equiv \frac{7 + \sqrt{48}\, \sqrt{1-\xi}}{1 + 48\,\xi} \,(1 + 6 \xi).
\end{align}
The function $F(\xi)$ \emph{monotonously decreases} for ${\xi \in [0,1]}$, with minimum ${F(1) = 1}$, so that ${F(\xi) \ge 1}$. The value $\Lambda^+$ is thus not admitted.

The complementary value $\Lambda^-$ \emph{yields a possible solution}, but the ranges of the metric parameters are restricted. Let us define dimensionless constants
\begin{align}\label{eq:DefPsi}
\psi \equiv \frac{e^2 + g^2}{a^2} \ge 0, \qquad\qquad
\eta \equiv 7 + 6\, \psi,
\end{align}
so that
\begin{align} \label{eq:ExpLambdaMinus}
\Lambda^- (\tfrac{1}{3}a^2 + 2l^2) = \big[\,\eta - \sqrt{\eta^2 - (1 + 48\,\xi)}\,\big]\,
 \frac{1 + 6\,\xi}{1 + 48\,\xi}.
\end{align}
This expression is required to be ${<1}$, which implies a constraint on the possible values of $l$. Using ${\eta \ge 7}$, we get ${6\,\xi < \eta - 5 + \sqrt{\eta^2 - 8\,\eta + 23}}$, that is
\begin{align}
l^2 < \tfrac{1}{3}a^2 + e^2 + g^2 + \sqrt{\tfrac{4}{9}a^4 + a^2 (e^2 + g^2) + (e^2 + g^2)^2}. \label{eq:UpperBoundForl}
\end{align}
For ${e=0=g}$ we simply obtain ${|l|<|a|}$ as in the previous case ${\Lambda<0}$.

Using the parameters introduced in \eqref{eq:DefF} and \eqref{eq:DefPsi}, the condition \eqref{eq:Lambda-pm-real>0}, which guaranties that $\Lambda^-$ is well-defined, can be rewritten in the form
\begin{align}
48 \xi < \eta^2 - 1.
\end{align}
Then it is easy to show that for all ${\eta \geq 7}$
\begin{align}
8(\eta - 5 + \sqrt{\eta^2 - 8\,\eta + 23}) \leq \eta^2 - 1.
\end{align}
The condition \eqref{eq:UpperBoundForl} thus \emph{restricts} the values of $l$ \emph{more} than \eqref{eq:Lambda-pm-real>0}.

Under the condition \eqref{eq:UpperBoundForl}, \emph{the multiple degenerate horizon is located at}
\begin{align} \label{eq:rH=rC}
r_H^2 = r_C^2 = \frac{1}{2\Lambda^-} - \tfrac{1}{6} a^2 - l^2,
\end{align}
where $\Lambda^-$ is given by \eqref{eq:LambdaPlusMinus}.

\subsection*{The case ${\Lambda > 0}$ with ${\Lambda \neq \Lambda^\pm}$}

In this general case with positive cosmological constant there exist \emph{two distinct extremal horizons at ${r_h^2\equiv x}$ if and only if}~${\Lambda \in (0,\Lambda^-)\cup(\Lambda^+, \infty)}$. Otherwise there are no horizons and the singularity is naked.

The roots $x_{\pm}$, explicitly given by \eqref{eq:x_pmAppendix}, must be positive.
The condition ${x_+ > 0}$ requires ${-q>-\sqrt{D}}$, which is equivalent either to ${q<0}$ or to ${s<0}$.
On the other hand, for ${x_- > 0}$ one needs ${-q>\sqrt{D}}$ which is ${q<0}$ and ${s>0}$. The latter conditions are stronger than the former, thus ${x_- > 0}$ implies ${x_+ > 0}$.

Let us investigate the condition ${x_- > 0}$. It differs from the ${\Lambda<0}$ case, because $q$ might be positive or negative as well, which induces \emph{an additional} constrain for $\Lambda$, not only for $l$. We require
\begin{align}
1)& \quad -q =  1 - \Lambda (\tfrac{1}{3}a^2 + 2l^2) > 0, \label{eq:cond-q}\\
2)& \quad  s = (a^2 - l^2)(1 - \Lambda\, l^2) + e^2 + g^2 > 0. \label{eq:cond s}
\end{align}
The first condition is \emph{violated by every} ${\Lambda > \Lambda^+}$ because
\begin{align}
\Lambda (\tfrac{1}{3}a^2 + 2 l^2) > \Lambda^+ (\tfrac{1}{3}a^2 + 2 l^2)  \geq  F(\xi) \ge 1,
\end{align}
where we used our previous estimate \eqref{eq:EstimateForLamda}. On the other hand, it is \emph{fulfilled by every} $\Lambda < \Lambda^-$ provided $l$ is bounded by \eqref{eq:UpperBoundForl}.

It is useful to introduce another dimensionless (positive) parameter
\begin{align}
\lambda \equiv a^2 \Lambda,
\end{align}
and analogously,
\begin{align}\label{eq:lambda+-}
\lambda^{-} \equiv a^2 \Lambda^- , \qquad\qquad
\lambda^{+} \equiv a^2 \Lambda^+.
\end{align}
The inequalities \eqref{eq:cond-q}, \eqref{eq:cond s} are then recast into the form
\begin{align}
1)& \quad 1 - \lambda\, (\tfrac{1}{3} + 2 \,\xi) > 0, \label{eq:cond-1}\\
2)& \quad 1 - \xi + \psi  - \lambda \, \xi (1-\xi) \ge 0. \label{eq:cond-2}
\end{align}

\begin{figure}[tb]
\begin{center}
\includegraphics[width=0.6\textwidth]{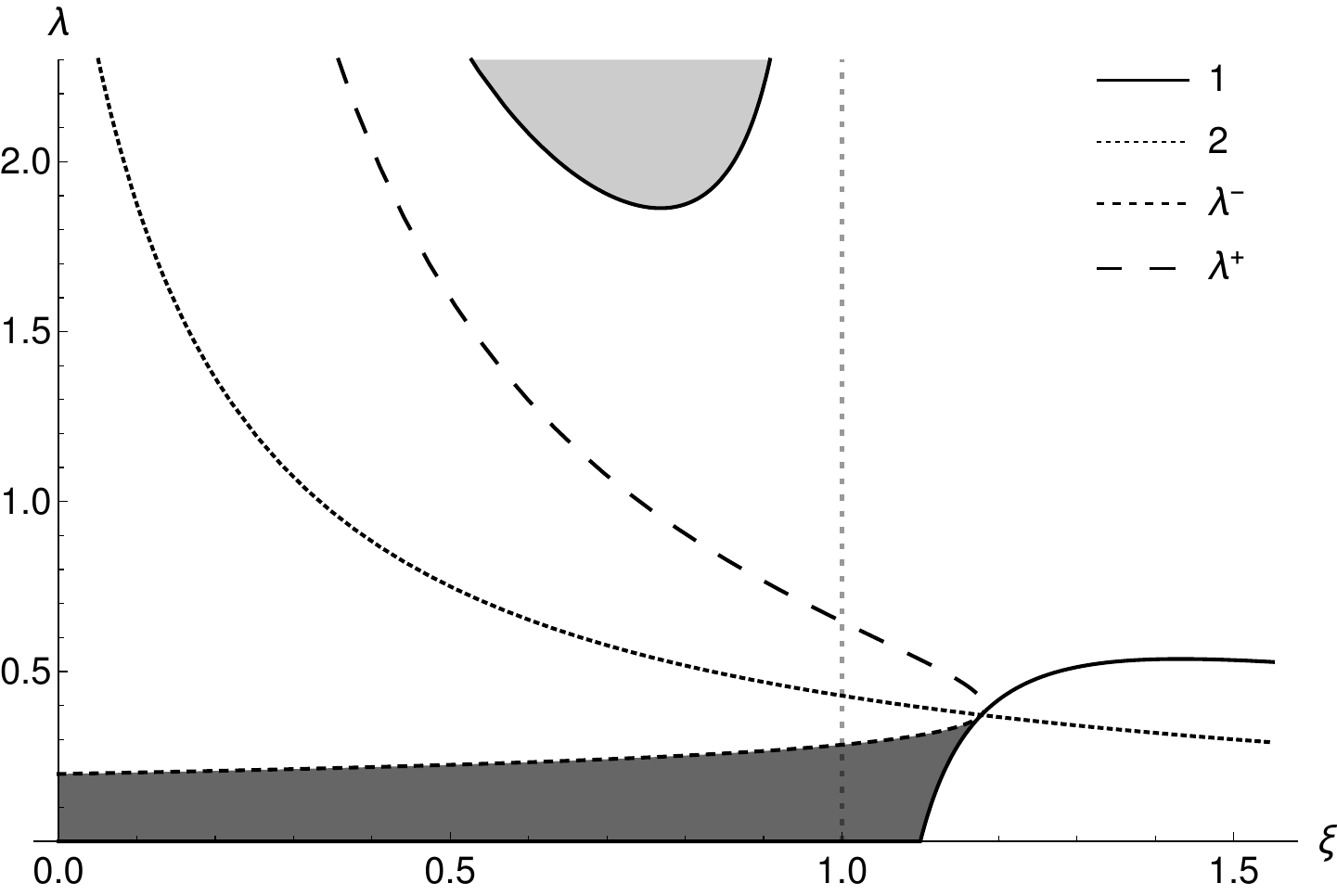}
\caption{Specific constraints on the values of the dimensionless metric parameters  ${\xi=l^2/a^2}$ and ${\lambda=a^2\Lambda}$. The inequality~\eqref{eq:InequalitiesForSmallLambda-1} is represented here as the curve~1, while the inequality~\eqref{eq:InequalitiesForSmallLambda-2} is represented as the curve~2. Also shown are the values $\lambda^{-} $ and $\lambda^{+}$ given by \eqref{eq:lambda+-}, \eqref{eq:LambdaPlusMinus}. The shaded areas denote the admissible values of the  parameters~$\xi$ and~$\lambda$. The dimensionless ``charge'' parameter $\psi$ was chosen here as ${\psi =0.1}$, but qualitative behaviour is the same for any ${\psi >0}$.} \label{fig:Inequalities}
\end{center}
\end{figure}

\emph{When} ${\xi < 1}$, that is for ${|l|<|a|}$, we may write these conditions as
\begin{align}
1)& \quad \frac{3}{1 + 6\, \xi} > \lambda, \label{eq:InequalitiesForSmallLambda-1}\\
2)& \quad  \frac{1 - \xi + \psi }{\xi (1-\xi)} > \lambda, \label{eq:InequalitiesForSmallLambda-2}
\end{align}
where the expressions on the left-hand sides are functions of $\xi$, also depending on the parameter $\psi$. For any fixed $\xi$ they determine the maximal value of $\lambda$. Admissible values of $\lambda$ for each $\xi$ are represented graphically by the \emph{dark shaded area} in Fig.~\ref{fig:Inequalities}.

For ${0<\xi<1}$ we have an estimate
\begin{align}
\frac{1 - \xi + \psi }{\xi (1-\xi)} > \frac{3}{1 + 6\, \xi} > \lambda^{-}
\end{align}
for any value of ${\psi \geq0}$. The cosmological constant is thus \emph{not} additionally restricted, and it remains ${\Lambda \in (0,\Lambda^-)}$.

\emph{For} ${\xi>1}$ we have to reverse the inequality \eqref{eq:InequalitiesForSmallLambda-2} which, apart from the upper bound $\lambda^-$, bounds the value of $\lambda$ from below
\begin{align}
\frac{\xi - 1 - \psi }{\xi (\xi - 1)} < \lambda < \lambda^- < \frac{3}{1 + 6\,\xi}.
\end{align}
In this case necessarily ${\psi>0}$. These inequalities imply a maximal value for $\xi$ as well, namely
\begin{align}
\xi < \tfrac{1}{3} + \psi + \sqrt{\tfrac{4}{9} + \psi(\psi+1)}.
\end{align}
Written in terms of the physical parameters, it is exactly the condition \eqref{eq:UpperBoundForl}. In particular, ${\psi=0}$ requires ${\xi<1}$, that is ${e=0=g}$ requires ${|l|<|a|}$.

To sum up, in the case of a positive cosmological constant ${\Lambda \in (0,\Lambda^-)}$ there is \emph{the extremal black hole horizon as well as the cosmological horizon}, located at
\begin{align}
r_H^2 &= \frac{1}{2\Lambda}\Big[1 - \Lambda\,(\tfrac{1}{3}a^2+2l^2) - \sqrt{D}\,\Big], \label{eq:rHAndrC}\\
r_C^2 &= \frac{1}{2\Lambda}\Big[1 - \Lambda\,(\tfrac{1}{3}a^2+2l^2) + \sqrt{D}\,\Big], \label{eq:rHAndrD}
\end{align}
provided the value of the NUT parameter $l$ satisfies the condition \eqref{eq:UpperBoundForl}. If ${|a|>|l|}$ the value of $\Lambda$ is not further restricted, while if ${|l|>|a|}$ and ${e^2 + g^2\neq0}$ there is a \emph{lower bound} for $\Lambda$ given by
\begin{align} \label{eq:LowerLambda}
\Lambda > \Lambda_0 \equiv \frac{a^2 - l^2 + e^2 + g^2}{ l^2\, (a^2 - l^2)} .
\end{align}

Finally, let us look at the second condition ${x_+>0}$. We have already shown that if ${q<0}$ then it is sufficient to have positive $x_+$ irrespective of the sign of $s$. However, if ${q>0}$ one can still ensure that $x_+$ is positive by requiring ${s<0}$. In such a case there is \emph{only a cosmological} horizon. In the interval ${0<\xi<1}$ (that is for ${|l|<|a|}$) we obtain the following restriction on the values of the cosmological constant
\begin{align} \label{eq:LambdaGreaterThanLambda0}
 \lambda^{+} < \frac{1 - \xi + \psi  }{\xi (1-\xi)} < \lambda.
\end{align}
In terms of the physical parameters it reads
\begin{align}
\Lambda^+ < \Lambda_0 < \Lambda.
\end{align}
Hence, the cosmological horizon exists for all values $\Lambda \in (\Lambda_0, \infty)$. However, if $\xi>1$ there is no positive solution, i.e.\ \emph{no horizons}. The values of $\lambda$, which satisfy equation \eqref{eq:LambdaGreaterThanLambda0} are graphically represented by the \emph{light shaded area} in Fig.~\ref{fig:Inequalities}.

Interestingly, the presence of the cosmological horizon depends not only on the value of the cosmological constant~$\Lambda$ but also on all other parameters of the black hole, namely of the mutual relation of the Kerr-like rotation~$a$ and the NUT parameter~$l$.


\end{document}